\documentclass{article}
\usepackage{amsfonts}
\usepackage{amsmath}

\newtheorem{theorem}{Theorem}

\newtheorem{example}[theorem]{Example}

\newtheorem{proposition}[theorem]{Proposition}
\newtheorem{remark}[theorem]{Remark}

\newenvironment{proof}[1][Proof]{\noindent\textbf{#1.} }{\ \rule{0.5em}{0.5em}}
\input{tcilatex}

\begin{document}

\title{HEAT KERNEL-ZETA FUNCTION RELATIONSHIP COMING FROM THE CLASSICAL
MOMENT PROBLEM.}
\author{M.Tierz \\
Laboratoire de Physique Th\'{e}orique et Mod\`{e}les Statistiques. \\
Universite Paris-Sud, B\^{a}timent 100, 91405 Orsay Cedex, France. \\
\&\\
Institut \ d'Estudis Espacials de Catalunya \\
(IEEC/CSIC). Edifici Nexus, Gran Capit\`{a} 2-4, 08034 \\
Barcelona, Spain.\\
tierz@ipno.in2p3.fr \& tierz@ieec.fcr.es \and E.Elizalde \\
Instituto de Ciencias del Espacio (CSIC) \& Institut \\
d'Estudis Espacials de Catalunya (IEEC/CSIC).\\
Edifici Nexus, Gran Capit\`{a} 2-4, 08034 Barcelona,\\
Spain. elizalde@ieec.fcr.es }
\maketitle

\begin{abstract}
By using ideas and strong results borrowed from the classical moment
problem, we show how ---under very general conditions--- a discrete number
of values of the spectral zeta function (associated generically with a
non-decreasing sequence of numbers, and not necessarily with an operator)
yields all the moments corresponding to the density of states, as well as
those of the partition function of the sequence (the two basic quantities
that are always considered in a quantum mechanical context). This goes
beyond the well known expression of the small$-t$ asymptotic expansion of
the heat kernel of an operator in terms of zeta function values. The precise
result for a given situation depends dramatically on the singularity
structure of the zeta function. The different specific situations that can
appear are discussed in detail, using seminal results from the zeta function
literature. Attention is paid to formulations involving zeta functions with
a non-standard pole structure (as those arising in noncommutative theories
and others). Finally, some misuses of the classical moment problem are
pointed out.
\end{abstract}

\section{Introduction}

There are two well known spectral functions associated with a sequence of
numbers $\{e_{n}\}$ (ordered, non-negative), which in a physical context may
correspond to some realistic quantum sequence. By realistic we mean than the
energy levels of a quantum system usually have some constraints in its
growth, given by Schr\"{o}dinger's equation. These two functions are the
following.

The trace of the heat kernel, or partition function: 
\begin{equation}
K(t)=\sum _{n=0}^{\infty }e^{-te_{n}},  \label{1}
\end{equation}
and the spectral zeta function 
\begin{equation}
\zeta (s)=\sum _{n=0}^{\infty }e_{n}^{-s}.  \label{2}
\end{equation}
By simple contour integration it is immediate to show that these two
spectral functions are related by a Mellin transformation (up to a gamma
function). The relation is: 
\begin{equation}
\zeta (s)=\frac{1}{\Gamma (s)}\int_{0}^{\infty }t^{s-1}\,K(t)\,dt=\frac{1}{%
\Gamma (s)}\,M[K(t);s]  \label{3}
\end{equation}
It follows from its definition that the Mellin transformed, $M[k(t);s]$,
gives all the moments of the function $K(t)$. In probabilistic terms, it
provides all the moments of a random variable whose probability density
function is proportional to $K(t)$. As we will see below, provided some
conditions are satisfied, the proportionality constant that normalizes $K(t)$
(as a probability measure) is given by $\zeta (1)^{-1}$. The extreme
importance of the possibility of normalization of the probability measure
will become clear below.

Let us also recall that, in the theory of pseudodifferential operators ($%
\Psi $DO) the relation between heat kernel and zeta functions is much more
profound. Let $A$ a $\Psi $DO, fulfilling the conditions of existence of a
heat kernel and a zeta function (see, e.g., \cite{Eli1}). Its corresponding
heat kernel is given by (see, for instance,\cite{Eli2}, and references
therein) 
\begin{equation}
K_{A}(t)=\mbox{Tr }e^{-tA}={\sum_{\lambda \in \mbox{Spec }A}}^{\prime
}e^{-t\lambda },
\end{equation}
which converges for $t>0$, and where the prime means that the kernel of the
operator has been projected out before computing the trace, and the
corresponding zeta function (as a Mellin transform) 
\begin{equation}
\zeta _{A}(s)=\frac{1}{\Gamma (s)}\mbox{Tr }\int_{0}^{\infty
}t^{s-1}\,e^{-tA}\,dt.
\end{equation}
For $t\downarrow 0$, we have the following asymptotic expansion: 
\begin{equation}
K_{A}(t)\sim \alpha _{n}(A)+\sum_{n\neq j\geq 0}\alpha
_{j}(A)\,t^{-s_{j}}+\sum_{k\geq 1}\beta _{k}(A)\,t^{k}\ln t,\quad
t\downarrow 0,
\end{equation}
being 
\begin{eqnarray}
&&\alpha _{n}(A)=\zeta _{A}(0),\quad \alpha _{j}(A)=\Gamma (s_{j})\mbox{Res }%
_{s=s_{j}}\zeta _{A}(s),\ \ \mbox{if }s_{j}\notin Z\ \mbox{or }s_{j}>0, 
\notag \\
&&\alpha _{j}(A)=\frac{(-1)^{k}}{k!}\left[ \mbox{PP } \zeta _{A}(-k)+\left(
1+\frac{1}{2}+\cdots +\frac{1}{k}-\gamma \right) \ \mbox{Res }_{s=-k}\zeta
_{A}(s)\right] ,  \notag \\
&&\beta _{k}(A)=\frac{(-1)^{k+1}}{k!}\,\mbox{Res }_{s=-k}\zeta _{A}(s), 
\notag
\end{eqnarray}
where PP means the principal part: 
\begin{equation}
\mbox{PP } \varphi =\lim_{s\to p}\left[ \varphi (s)-\frac{\mbox{Res }%
_{s=p}\varphi (s)}{s-p}\right] ,
\end{equation}
a finite number. It turns out that all $\beta _{k}=0$ if $A$ is a
differential operator. Note that, in the case of a generic $\Psi $DO
fulfilling the conditions of existence of the zeta function, this one
provides the asymptotic expansion ($t\downarrow 0$) of the heat kernel of
the operator.

The main result we are going to obtain here can be viewed as a sort of
complement or completion of the above expansion. Rather than dealing with
the asymptotics of the heat kernel we will make use of results known for the
classical moment problem in order to gain knowledge of the heat kernel
itself, from the behavior of the zeta function. And this, moreover, will not
be restricted to a family of differentiable or $\Psi $DOs, but will hold in
a quite general setting.

Now we comment briefly on some basic issues of the moment problem (for a
recent review and further details see {\cite{Simon}}). The two fundamental
moment problems are the following.

\subsubsection*{\textbf{Problem 1.} The Hamburger moment problem.}

Given a sequence of reals $E_{0},E_{1},... $, does it exist a measure, $%
d\rho $, on $(-\infty ,\infty )$, so that: 
\begin{equation*}
E_{n}=\int _{-\infty }^{\infty }x^{n}d\rho (x)
\end{equation*}
and, if such a measure exists, is it unique ?

\subsubsection*{\textbf{Problem 2.} The Stieltjes moment problem.}

Given a sequence of reals $E_{0},E_{1},... $, does it exist a measure, $%
d\rho $, on $(0,\infty )$, so that: 
\begin{equation*}
E_{n}=\int _{0}^{\infty }x^{n}d\rho (x)
\end{equation*}
and, if such a measure exists, is it unique ?

There is an extensive theory behind these so simply stated problems. In our
case, we will be interested in just a couple of results, that we quote
explicitly here because of their simplicity, and also in order to keep this
paper self-contained. In this context, the following proposition is useful.

\begin{proposition}
Suppose that $\left\{ E_{n}\right\} _{n=0}^{\infty }$ is a set of Hamburger
moments and that for some $C,R>0,$%
\begin{equation*}
\left| E_{n}\right| \leq CR^{n}n!,
\end{equation*}
Then the Hamburger problem is determinate. If $\left\{ E_{n}\right\}
_{n=0}^{\infty }$ is a set of Stieltjes moments and 
\begin{equation*}
\left| E_{n}\right| \leq CR^{n}(2n)!.
\end{equation*}
Then the Stieltjes moment problem is determinate.
\end{proposition}

\section*{2. Encoding of the heat kernel in terms of zeta values}

The preceding proposition is quite interesting, since it provides a
sufficient condition for the existence and uniqueness of the solution to the
problems. Also, it follows from equation \ref{3} that $\zeta
(s)=E_{s-1}/\Gamma (s)$, for integer s, and the (non-normalized) measure is
given by $d\rho (t)=K(t)dt$.

That is, we have that the moments, relative to our measure, are given by: 
\begin{equation}  \label{4}
E_{n}=\zeta (n+1)\Gamma (n+1)=\zeta (n+1)\, n!.
\end{equation}
As we can see, this is a remarkable relationship, because the connection
between the two spectral functions ---heat kernel and zeta function--- stems
not only from the fact that the second is the Mellin transformed of the
first, and that then it encodes all the moments of the first, but especially
that it is the Mellin transform up to a gamma function, so it is, in some
sense, even more tied to the problem of moments since it introduces the
factorial. And this implies that it is not only a Stieltjes determined
problem but also that we can construct an associated Hamburger problem that
is also determinate, since it saturates the bound. On the other hand, the
analytical continuations of the zeta and gamma functions on the variable $n$
to the whole of the complex plane, provides also a natural extension of the
corresponding moments to complex values of the argument.

We can now state our main result.

\begin{theorem}
The trace of the heat kernel $K(t)$ is a function exactly reproducible by
its moments only if its associated zeta function $\zeta (s)$ does not have
any pole at any integer value of $s.$
\end{theorem}

\begin{proof}
From equation \ref{4} we see that the moments of the heat kernel satisfy the
bound stated in the proposition, as long as $\zeta (n+1)\leq CR^{n}$, which
is certainly the case if there is no singularity in the zeta function for
any positive integer $n$.
\end{proof}

\begin{example}
From the well known pole at $s=1$, that we can interprete as lack of
normalization of the associated heat kernel, we should immediately exclude
from the list of the functions that satisfy the theorem the following
important cases.

1. The Riemann zeta function: $\zeta \left( s\right) =\sum_{n=1}^{\infty
}n^{-s}.$

2. The Hurwitz zeta function: $\zeta (s,a)=\sum_{n=0}^{\infty }\left(
n+a\right) ^{-s}$

3. The Epstein zeta function in dimension two: 
\begin{equation*}
E\left( s;a,b,c\right) =\sum_{n,m}\left( am^{2}+bmn+cn^{2}\right) ^{-s}
\end{equation*}
$.$

All of them having its rightmost pole at $s=1$. Trivially, all these zeta
functions give rise to indeterminate problems, in spite of the fact that all
of their moments, except the zeroth one, are perfectly well behaved. Their
associated heat kernels are not normalizable. One example of well-behaved
(in our context) zeta function is: 
\begin{equation*}
G\left( s;a,c;q\right) =\left[ a\left( n+c\right) ^{2}+q\right] ^{-s}.
\end{equation*}
It is often called Epstein-Hurwitz in the physics literature \cite{Eli1}\cite%
{Eli3}, and it has its rightmost pole at $s=\frac{1}{2}$.
\end{example}

Let us now quote another Proposition from {\cite{Simon}} that will help us
to check some of the statements mentioned above, and to get some identities.

\begin{proposition}
(Krein \cite{Simon}) Suppose that $d\rho (t)=K(t)dt,$ where $0\leq K(t)\leq
1 $ and either

$\left( i\right) $ $\sup p\left( K\right) =\left( -\infty ,\infty \right) $
and 
\begin{equation*}
\int_{-\infty }^{\infty }\frac{\ln K(t)dt}{1+t^{2}}<\infty ,
\end{equation*}

$\left( ii\right) $ $\sup p\left( K\right) =\left( 0,\infty \right) $ and 
\begin{equation*}
\int_{0}^{\infty }\frac{\ln K(t)dt}{1+t}\frac{1}{\sqrt{t}}<\infty ,
\end{equation*}
Suppose also that, for all $n:$%
\begin{equation*}
\int_{-\infty }^{\infty }\left| t\right| ^{n}K(t)<\infty ,
\end{equation*}

Then the moment problem, with moments: 
\begin{equation*}
E_{n}=\frac{\int t^{n}K(t)dt}{K(t)dt},
\end{equation*}
is indeterminate.
\end{proposition}

\begin{remark}
By the same method, Krein proved a stronger result: $K(t)$ need not be
bounded and the measure defining the moments can have an arbitrary singular
part. It is also important to comment that Krein's conditions are close to
optimal. That is, for example, if the moments under consideration saturate
the conditions stated in our first proposition, then the integrals in
Krein's proposition are barely divergent. Thus, taking into account our
theorem and the last proposition we have the following one.
\end{remark}

\begin{proposition}
For the kind of heat kernels that belong to the category stated in the
theorem, we have that the following integral: 
\begin{equation*}
\int_{0}^{\infty }\frac{\ln K(t)dt}{t^{1/2}+t^{3/2}},
\end{equation*}
is divergent.
\end{proposition}

\section*{3. The example of the theta function.}

\begin{example}
Let us consider the family of functions $K(t)=\sum_{n=1}^{\infty
}e^{-n^{\alpha }t},$ with $\alpha $ a real parameter with $\alpha \neq \frac{%
1}{n}$ with $n$ any integer (thought it is maybe necessary to distinguish
between $\alpha >1$ and $\alpha <1$). Its associated zeta function has the
rightmost pole moved from $s=1$ in the Riemann case $\left( \alpha =1\right) 
$ to $s=\frac{1}{\alpha }$. Thus, in principle this is an example of the
kind of function that satisfies our theorem and also the last proposition.
\end{example}

Let us apply what we have learned up to now to a very basic situation. From
what we have said, it is clear that we cannot work out the Riemann case,
since it has a pole at $s=1$. But it turns out that we can in fact consider
the following heat kernel: 
\begin{equation}
K(t)=\sum_{n=1}^{\infty }e^{-n^{2}t},
\end{equation}
which is a particular case of the elliptic theta function: 
\begin{equation}
\theta _{3}(z,\tau )=\sum_{n=-\infty }^{\infty }e^{-\pi n^{2}\tau +2\pi nz}
\end{equation}
with $z\in C,\tau \in R^{+}.$ More precisely: $K(t)=\frac{1}{2}[\theta
_{3}(0,t/\pi )-1]$

Thus, we are dealing with a particular case of a theta function. Its zeta
function counterpart is: $\widetilde{\zeta }(s)=\sum _{n=1}^{\infty
}n^{-2s}=\zeta _{R}(2s) $, where the last one is the Riemann zeta function.

It is well known that with theta functions one is usually able to obtain
remarkable identities, some of them coming basically from the Jacobi
identity (or the ``equivalent'' summation formulas), but not so often
completely closed, evaluable expressions. In the light of the theory of
moments and from what has been stated above, we now have all the information
about the theta function encoded in the integer values of $\zeta _{R}(2s)$.
But the theory of moments does not say anything about how to do the
reconstruction. We can try, for example, the well known ``probabilistic
approach'', that consists in the construction of the characteristic or
generating function. In general, the characteristic (or generating) function
is given by: 
\begin{equation}
\phi (k)=\sum_{n=0}^{\infty }\frac{<X^{n}>}{n!}k^{n}=\sum_{n=1}^{\infty }%
\frac{\zeta (n)\Gamma (n)}{(n-1)!}k^{n-1}=\sum_{n=1}^{\infty }\zeta
(n)k^{n-1}.
\end{equation}
And, in the particular case we are studying, 
\begin{equation}
\phi (k)=\sum_{s=1}^{\infty }\zeta _{R}(2s)(-k)^{s}=\frac{-1+\pi \sqrt{k}%
\coth (\pi \sqrt{k})}{2k}.
\end{equation}
In principle, this expression should be equal to the Laplace transform of
our theta function, and this would imply a confirmation of our theorem in
this particular case. Let us show that this is indeed the case. Consider 
\begin{equation}
\int_{0}^{\infty }e^{-sx}\sum_{n=1}^{\infty
}e^{-n^{2}x}dx=\sum_{n=1}^{\infty }\frac{1}{n^{2}+s},
\end{equation}
and using the well known series expansion of the cotangent, we get that the
Laplace transform is just our $\phi (k)$. Thus, such very simple case
provides a practical confirmation of our theorem. Now we have two possible
different ways to arrive to the same result. The direct one, of computing
the Laplace (or Fourier) transform of the theta function, and the other ---a
consequence of our theorem--- by a direct construction using the associated
zeta function. It would be interesting to investigate which kind of
information can be obtained from the comparison of the two methods in more
complicated or practical cases than the one considered here. On the other
side, we see that the Laplace transform of our theta function presents
periodic singularities at $k=n^{2}$, with $n=1,2,3...$ It would be
interesting to know whether this can be related with to fact that our theta
function, for example, has some kind of special features like the lack of an
analytical continuation around the origin. Or, more probably, to the
presence of the pole at $s=\frac{1}{2}$ of the associated zeta function, to
which the theory of moments is blind (and that always implies the divergence
of a certain integral associated to the heat kernel). In this particular
case, the divergent integral is $\int_{0}^{\infty }K(t)/\sqrt{t}$.

\section*{4. Inversion problems and possible encoding of the heat kernel in
terms of negative zeta values}

The classical moment problem, as can be readily appreciated from \cite{Simon}%
, has to do essentially with the uniqueness issue. In this spirit, we have
seen that, under the restriction on the position on the poles, we can
guarantee determinacy: the positive integer values of the zeta function
determine in a unique way the partition function of the system. This is
equivalent to saying that there is only one partition function with the same
given moments: if there is another with the same integer values of the
associated zeta function it should necessarily be the same partition
function. This is an important formal statement, since it implies that only
with a very restricted part of the information included in the spectral zeta
function (the set of positive integer values against the whole complex
plane), we have completely determined and fixed the partition function.
Nevertheless, it is not guaranteed that the inversion, from the practical
point of view, can always be performed in a straightforward way, as
explained in the previous section. More precisely, the construction of the
characteristic function used above is based on the following series and
integral commutation : 
\begin{equation*}
\int_{0}^{\infty }e^{-\beta t}K(t)dt=\int_{0}^{\infty }\sum_{k=0}^{\infty }%
\frac{\left( -\beta t\right) ^{k}}{k!}K(t)dt=\sum_{k=0}^{\infty }\left(
\int_{0}^{\infty }t^{n}K(t)dt\right) \frac{(-\beta )^{k}}{k!}
\end{equation*}
\begin{equation}
=\sum_{k=0}^{\infty }M[f(t);k]\frac{\left( -\beta \right) ^{k}}{k!}%
=\sum_{k=0}^{\infty }\zeta (k+1)(-\beta )^{k}.
\end{equation}
Of course, this is cannot b always carried out in such a simple way (see 
\cite{Asympt}, and \cite{QFT3} for a discussion in a QFT context). In our
case, our main concern will be precisely the situation when regularized
values appear (see \cite{Eli1}). As we will see later, it is of course also
possible to do this in such cases, since we deal with a function $K(t)$ of
the type $K(t)=\sum_{n}e^{-e_{n}t}$. Then we can proceed as 
\begin{equation}
\int_{0}^{\infty }e^{-\beta t}K(t)dt=\int_{0}^{\infty }e^{-\beta
x}\sum_{n=1}^{\infty }e^{-e_{n}x}dx=\sum_{n=1}^{\infty }\frac{1}{e_{n}+\beta 
}=\sum_{n=1}^{\infty }\sum_{k=0}^{\infty }\frac{(-\beta )^{k}}{e_{n}^{k+1}}.
\end{equation}

Clearly, the two expression are only equal under the validity of the
commutation of the two series in the last expression: 
\begin{equation}
\sum_{n=1}^{\infty }\sum_{k=0}^{\infty }\frac{(-\beta )^{k}}{e_{n}^{k+1}}%
=\sum_{k=0}^{\infty }\sum_{n=1}^{\infty }\frac{(-\beta )^{k}}{e_{n}^{k+1}}%
=\sum_{k=0}^{\infty }\zeta (k+1)(-\beta )^{k}.
\end{equation}
This is exactly the case of the example considered, as checked above, and it
is to be expected in general whenever the rightmost pole is at $s<1$ and, in
addition, it has been shown that the moments, as in our case, directly
satisfy the bound that guarantees that the Taylor expansion can be done.
Nevertheless, the case with $s>1$ should be considered with more care since
regularized values for the moments appear and it is in general known that
commutation of series cannot be performed in a direct way, as we will see.
To begin with, a representation of the Laplace transform by $%
\sum_{n=1}^{\infty }\frac{1}{e_{n}+\beta }$ is clearly a regularized series
if the rightmost pole is at the right of $s=1$.

We now consider the possibility that the zeta function encodes, in the
positive integer values of its argument, all the information that describes
the heat kernel (and not only that of its asymptotic expansion). Again, from
the very simple expansion of the exponential: $e^{-e_{n}t}=\sum_{j=0}^{%
\infty }\frac{(-e_{n}t)^{j}}{j!}$, we can wonder whether 
\begin{equation}
K(t)=\sum_{n=1}^{\infty }\sum_{j=0}^{\infty }\frac{-(e_{n}t)^{j}}{j!}
\end{equation}
has something to do with the following quantity: 
\begin{equation}
\widetilde{K}(t)=\sum_{j=0}^{\infty }\sum_{n=1}^{\infty }\frac{(e_{n})^{j}}{%
j!}t^{j}=\sum_{j=0}^{\infty }\frac{\zeta (-j)(-t)^{j}}{j!}.
\end{equation}

The first thing one can wonder about is, whether these two expressions are
equal, that is, if we can proceed with the commutation of the series. The
validity of the commutation of the two series would imply the use of the
regularized sums $\sum_{n}e_{n}^{j}$ (essentially this is a divergent series
and we use its regularized value, expressed by the zeta function). So, this
makes unlikely that this is the case, and actually, it has been shown in the
theory of spectral zeta functions that commutations of series cannot be
done, in general, without the additional contributions {\cite{Eli1}}. This
implies, in principle, that we cannot reproduce the heat-kernel from the
values of the associated zeta function at negative integer values of $s$, at
least not in such a naive way. Nevertheless, we can use the functional
equation satisfied by the zeta function.

In the particular example considered, it is the well known functional
equation for the Riemann zeta function {\cite{Titch}}: 
\begin{equation}
\pi ^{-s/2}\Gamma (s/2)\zeta _{R}(s)=\pi ^{-(1-s)/2}\Gamma (\frac{1-s}{2}%
)\zeta _{R}(1-s)
\end{equation}
It provides the analytic continuation of the Riemann zeta function to the
whole complex plane (with the pole at $s=1$), so it allows to express the
negative integer values of the zeta function in terms of the positive ones.
Then, we arrive at the formal result that the existence of a functional
equation implies that our statement also holds for the negative integer
values of the zeta function (but then we probably have to include gamma
functions and other factors in order to reproduce the numerical value of the
zeta function in the positive integer value, which actually is the
fundamental quantity in our study).

\section*{5. The role of the non-integer poles and the regularized values of
the moments}

A natural question that arises is: What is the role played by the pole or
poles at the non-integer positions in the determinate case ? To answer this
we consider the Laplace transform of a generic heat kernel: 
\begin{equation}
\int_{0}^{\infty }e^{-\beta x}\sum_{n=1}^{\infty
}e^{-e_{n}x}dx=\sum_{n=1}^{\infty }\frac{1}{e_{n}+\beta }
\end{equation}
We see that the Laplace transform is related to the associated spectral zeta
function in the following simple way. Being the associated spectral zeta
function $\zeta (s)=\sum e_{n}^{-s}$, we associate to this last one, a 
\textit{shifted} spectral zeta function: $\zeta (s,\beta )=\sum (e_{n}+\beta
)^{-s}$ (in particular, this leads from the Riemann zeta function to the
Hurwitz zeta function). We see clearly, that the Laplace transform of the
heat kernel is the shifted zeta function evaluated at $s=1$ and with the
parameter $\beta $ being the variable of the function. Since the addition of
the shift does not change the rightmost pole, we see that the divergence at $%
s=1$ is the most problematic one, since our expression for the Laplace
transform is expressed then as a genuine divergent series. This is in some
sense, to be expected, since the pole at $s=1$ implies lack of normalization
of the heat kernel. For the case when we have the rightmost pole at $s>1$,
the expression is a regularized series, while for $s<1$ it is a convergent
series.

Let us now study the case where $e_{n}=n^{\alpha }$. Then, the associated
zeta function has its rightmost pole at $s=\frac{1}{\alpha }$. Thus, for $%
\alpha $ running from $1$ to $\infty $, the rightmost pole moves from $s=1$
to $s=0$. And we see that the pattern of singularities in the Laplace
transform of the heat kernel goes like $n^{\alpha }$, so we have less and
less singular behavior corresponding to the pole tending towards the origin.
With this simple example we observe that, in spite of the fact that,
regarding uniqueness and inversion questions, only the positive integer
moments play a role (and this is due to the fact that the distribution and
its moments are related by a Taylor series expansion of a certain function
of the distribution), we see how the non-integer poles do also play a role,
not in the fact that the problem is determinate or not, but in the actual
form and behavior of the characteristic function (the integral transform of
the distribution). This was to be expected.

Now, consider that the parameter $\alpha $ ranges between $(0,1)$. Then we
move the rightmost pole to the right of $s=1$ (of course, avoiding values of
the parameter such that we get a pole in another integer position). We see,
directly from the Laplace transform, that the more the pole goes to the
right, the more divergences are to be accounted for. It is worth to remark
that in this case, all the zeta function values that are on the left of the
pole (that defines the abscissa of convergence) correspond to regularized
values, and this implies that in our series expansion we have a finite
number of moments whose expression is a regularized one. This is in contrast
with the case discussed above, and we can in fact relate it with the
commutation of series above mentioned.

\section*{6. The density of states in terms of the zeta function}

Let us now consider the density of states of the system: 
\begin{equation}
\rho (E)=\sum\limits_{n}\delta \left( E-e_{n}\right).
\end{equation}
Its relation with the zeta function can be simply written as: 
\begin{equation}
\zeta (s)=\sum_{n}e_{n}^{-s}=\int_{-\infty }^{\infty }\sum_{n}\delta
(E-e_{n})E^{-s}=\int_{-\infty }^{\infty }\rho (E)E^{-s}dE
\end{equation}
That is, the zeta function at its negative integer values gives all the
(regularized) moments of the density of states. As long as there are no
poles, this implies that the negative integer values of the zeta function
determine in a unique way the density of states, thanks to the functional
equation, that gives us the analytical prolongation to the whole complex
plane and implies, together with the considerations for the partition
function case, that the bound for the moments is equally satisfied.

As before, some difficulties can appear at the practical level, especially
when one is dealing with a certain number of moments (a finite number in the
previous example, with $\left( 0,1\right) $, and an infinite number in this
case) that come from a regularized divergent series. But, at least in this
last case, taking into account that the Laplace transform of the density of
states is the own heat kernel, and that the result on commutation on series
(or zeta function regularization theorem) tells us that: 
\begin{equation}
K(t)=\sum_{n}e^{-te_{n}}=\sum_{n}\sum_{j}\frac{(-t)^{j}e_{n}}{j!}\neq
\sum_{j}\sum_{n}\frac{(-t)^{j}e_{n}}{j!}=\sum_{j}\frac{(-\beta )^{j}}{j!}%
\zeta (-j).
\end{equation}
The crucial point is to understand that the classical moment problem give us
several conditions that lead us know whether the measure is unique or not.
There is a very simple and enlightening example, due to Stieltjes, and
included in, e.g. \cite{Simon}. It is based on the following observation.

For any $\theta \in \left[ -1,1\right] ,$ we have that 
\begin{equation}
\int_{0}^{\infty }u^{k}u^{-\ln u}\left[ 1+\theta \sin (2\pi \ln u)\right] =%
\sqrt{\pi }e^{\frac{1}{4}\left( k+1\right) ^{2}},
\end{equation}%
and we thus arrive to a one-parameter family of different functions all of
them having the same sequence of moments. Clearly, $\gamma _{k}=e^{\frac{1}{4%
}\left( k+1\right) ^{2}}$ is an indeterminate set of Stieltjes moments. It
is not necessary that the relationship with the moments should be given as
the usual series expansion in straightforward probability theory for
example. It is valid for the simple example of the theta function, shown
above. In fact there are papers in the physical literature dealing with the
classical moment problem for the case of the moments of a random quantity,
but they fail to appreciate altogether, that the condition on the bound we
have used to show determinacy in our context, is just a \textit{sufficient}
condition, that came from analyticity of the Fourier transform of the
measure. In fact, as clearly stated in \cite{Simon}, there can be sets of
moments of an arbitrary high growth which constitute nevertheless a
determinate moment problem. One should be careful since, on the other hand,
there are sets of moments that are only slightly bigger than the ones in our
proposition (1) and that constitute an indeterminate moment problem. Let us
go further into the results of the commutation of series in order to see
what happens with a practical implementation of the relationship between the
density of states and the trace of the heat kernel by the moments of the
last one. In \cite{Eli1}, it is shown that: 
\begin{equation}
K_{\alpha }(t=1)=\sum_{n=1}^{\infty }e^{-n^{\alpha }}=\sum_{a=1}^{\infty }%
\frac{(-1)^{a}}{a!}\zeta (-a\alpha )+\frac{1}{\alpha }\Gamma (\alpha
)-\Delta _{\alpha },
\end{equation}%
being $\Delta _{\alpha }$ an additional contribution coming from series
commutation. The particular case of the zeta function regularization says
that: 1. For $-\infty <\alpha <2$, the contribution of the additional term
is $\Delta _{\alpha }=0$. 2. For $\alpha =2,$ the contribution of the
additional term is $\Delta _{\alpha }=-\sqrt{\pi }K_{2}(\pi ^{2}).$ 3. For $%
\alpha >2,$ the contribution is increasingly bigger. Following our line of
reasoning, we identify each situation with the position of the rightmost
(and in this case only) pole, the correspondence being 1. The pole of the
zeta function lies in $(-\infty ,0)\cup \left( \frac{1}{2},\infty \right) .$
2. The pole of the zeta function is at $s=\frac{1}{2}.$ 3. The pole lies in $%
\left( 0,\frac{1}{2}\right) .$ No conclusion seems to be implied by this
observation. Notice, nevertheless, that we are in the particular case $t=1.$
For the general case, the expression: 
\begin{equation}
K_{\alpha }(t)=\sum_{n=1}^{\infty }e^{-n^{\alpha }t}=\sum_{a=1}^{\infty }%
\frac{(-t)^{a}}{a!}\zeta (-a\alpha )+\frac{1}{\alpha }\Gamma (\alpha
)-\Delta _{\alpha }
\end{equation}%
is only valid for $\alpha \in (0,1]$ and for $\alpha \in 2N.$ These last
value of the parameter are the ones that make the associated operator a
differential operator.

We have restricted our discussion up to now to zeta functions that only
posses one pole. It is worth to complete our characterization comparing with
cases where more poles appear. In particular, when we have an infinite
sequence of poles in the left half-plane. These appear in a natural way,
when we introduce a constant shift in the spectrum. In this case, we go, for
example, from the Hurwitz zeta function to the Epstein-Hurwitz zeta function
mentioned above: 
\begin{equation}
G(s;a,c;q)=\sum_{n=-\infty }^{\infty }[a(n+c)^{2}+q]^{-s}.
\end{equation}
This has, like the Hurwitz or the Riemann with $\alpha =2$, its rightmost
pole at $s=\frac{1}{2}$, but exhibits, in addition, an infinite number of
poles at $s=-\frac{1}{2},-\frac{3}{2},...$ produced by the $q$ term. The
Laplace transform of the heat kernel is: 
\begin{equation}
\pounds \{K(t)\}=\int_{0}^{\infty }e^{-\beta x}\sum_{n=1}^{\infty
}e^{-[a(n+c)^{2}+q]x}dx=\sum_{n=1}^{\infty }\frac{1}{[a(n+c)^{2}+q]+\beta }
\end{equation}
We see that the introduction of the parameter $q$, the responsible for the
new poles, only shifts by a constant value the pattern of divergences of
this function.

It should be clear that we can apply our results as soon as we know the
structure of the zeta function. This allows us to consider, for example,
cases as general as: 
\begin{equation}
\sum\limits_{n_{1},...n_{N}}[a_{1}(n_{1}+c)^{\alpha
_{1}}+...+a_{N}(n_{N}+c_{N})^{\alpha _{N}}+q]^{-s},
\end{equation}
with its rightmost pole at $s_{0}=\mbox{max }(\alpha _{1},\ldots ,\alpha
_{N})$. 
Notice that this example (considered in detail in the seminal paper \cite%
{Eli3}) does \textit{not} correspond generically to a $\Psi $DO. As we
pointed out before, our approach is not necessarily limited to $\Psi $DOs.
But, even in the family of $\Psi $DOs, very general pole structures for the
zeta function can show up (in particular, in the case of noncommutative
spaces, see e.g. \cite{Eli4,bez1,fpw1}).

\section*{7. The problem in the framework of noncommutative geometry}

It is clear that we can state the condition of absence of a pole at a
certain position in more mathematical terms as a zero value for the residue
of the zeta function at the point. This allows a more direct comparison with
usual heat kernel asymptotic expansion since, as we have seen, the
coefficients are given essentially by the residues of the zeta function at
certain points. To begin with, we have seen that, if the operator is a
differential operator then all $\beta _{k}=0$ and, since $\beta _{k}(A)=%
\frac{(-1)^{k+1}}{k!}\,\mbox{Res }_{s=-k}\zeta _{A}(s),$ then we have
automatically implemented the condition $\,\mbox{Res }_{s=-k}\zeta
_{A}(s)=0, $ that is, absence of pole at the negative integer values. Then
we can clearly state that for a differential operator the density of states
is uniquely determined by all its moments, that is, by the values of the
associated zeta function at its negative integer values. Of course, the
condition $\beta _{k}=0$ does not imply that the operator is differential,
and we have seen an example in the case of the spectrum $\left\{
e_{n}\right\} =n^{\alpha }.$ For, then we have always $\beta _{k}=0$, but it
corresponds to a differential operator only when $\alpha \in 2N.$ Also, for
example, we see that determinacy for the density of states not only implies $%
\beta _{k}=0,$ but also that the $\alpha _{j}(A)=\frac{(-1)^{k}}{k!}\zeta
\left( -k\right) $ (for the case $s=-k).$

We can now put our problem into the context of noncommutative geometry,
given that the tool used to compute the zeta function residues at any
position of the (hypothetical) pole is the Wodzicki residue \cite{Eli3}. The
Wodzicki (or noncommutative) residue \cite{NC1},\cite{NC2},\cite{NC3} is the
only extension of the Dixmier trace to $\Psi $DOs which are not in $\mathcal{%
L}^{(1,\infty )}$. Even more, it is the only trace at all one can define in
the algebra of $\Psi $DOs up to a multiplicative constant. It is explicitly
given by the integral 
\begin{equation}
\mbox{res}\ A=\int_{S^{*}M}\mbox{tr}\ a_{n}(x,\xi )\,d\xi ,
\end{equation}
with $S^{*}M\subset T^{*}M$ the co-sphere bundle on $M$ (some authors put a
coefficient in front of the integral). If dim $M=n=-$ ord $A$ ($M$ compact
Riemann, $A$ elliptic, $n\in \mbox{\bf N}$) it coincides with the Dixmier
trace, and one has: 
\begin{equation}
\mbox{Res}_{s=1}\zeta _{A}(s)=\frac{1}{n}\,\mbox{res}\ A^{-1}.
\end{equation}

An interesting property of the Wodzicki functional is that it is also the
Cauchy residue for the zeta function \cite{Eli3}. Then we can express our
condition of absence a pole in an integer position in the following simple
way: 
\begin{equation}
\int_{S^{*}M}\mbox{tr}\ a_{n}(j,\xi )\,d\xi =0,
\end{equation}
with $j=1,2,...$ for the uniqueness of the trace of the heat kernel and $%
j=0,-1,-2,...$ for the density of states. It is clear that a sufficient
condition to satisfy this requirement is that $a_{n}(j,\xi )=0$ for the
respective values of $j.$ Furthermore, in noncommutative geometry we define
a geometric space by a spectral triple \cite{NC1},\cite{NC2},\cite{NC3} : 
\begin{equation*}
(A,H,D),
\end{equation*}
where $A$ is a concrete algebra of coordinates represented on a Hilbert
space $H$ and the operator $D$ is the inverse of the line element. This is a
completely spectral definition, where the elements of the algebra are
operators and the points come from the joint spectrum of operators and the
line element is an operator. More precisely, in this framework the operator $%
D$ gives the inverse of the length element $ds=D^{-1}.$ The basic properties
of spectral triples are simple to state, and do not make any reference to
the commutativity of the algebra $A$ (see \cite{NC1},\cite{NC2},\cite{NC3}
for details). For our purposes, the following result for the case of
spectral triples is noticeable: 
\begin{equation*}
\int \left| D\right| ^{-n}\neq 0.
\end{equation*}
where $n$ is the dimension of the spectral triple. This amounts to saying
that the residue of the zeta function associated with the operator $D,$ $%
\zeta (s)=Tr\left( \left| D\right| ^{-s}\right) $ at $s=n$ cannot vanish.
Being $n$ an integer, we see clearly that this important case does not
fulfill our conditions leading to determinacy for the heat-kernel.

\section*{8. Conclusions and outlook}

The important result in this paper is the proof that the spectral zeta
function gives all the moments associated with the density of states as well
as those coming from the partition function, which are the two fundamental
quantities always studied in a quantum mechanical context. The moments of
the density of states are obtained from the zeta function at negative
integers and at the origin while zeta at the positive integers yields all
the moments of the partition function. Interestingly enough, this
observation provides also a clear physical motivation for the functional
equation satisfied by the spectral zeta function. Namely the functional
equation relates the moments of these two physical quantities, that are in
fact connected through a Laplace transform. This leads us, in fact, to a
functional equation of the type $\zeta (-s)=\varphi \left[ \Gamma (s+1)\zeta
(s+1)\right] $. The functional equation gives the analytical continuation of
the zeta function, and in this context it is very clear why it is important
to know it, not only in a quantum field theory context, but as a fundamental
object.

These considerations lead to an exact comparison with the classical moment
problem, to arrive at the important formal result that, under the natural
conditions stated, both the partition function as well as the density of
states are \textit{uniquely} determined by their respective moments: the
corresponding zeta values. As already emphasized in the paper, only with the
information encoded in the integer values of the zeta function we have
determined both the density of states as well as the partition function. In
the absence of the poles in the key positions, this is a clear and direct
consequence of the boundedness of the values of the zeta function (that are
in fact decreasing, due to the increasing nature of the spectral sequence)
and of the existence of a functional equation for the zeta function.

In addition, the consideration of inversion questions, related to practical
implementations of this result ---that lie usually beyond the scope of the
classical moment problem--- has lead us to interesting observations
regarding the commutation of series and the role and influence of the
non-integer poles, studied in the context of the integral transform of the
partition function.

\section*{Appendix A. The classical moment problem: some results and
considerations}

We have discussed essentially two different aspects of the classical moment
problem. The proposition, that gives a sufficient condition through a bound
on the sequence of moments, in order to prove determinacy, and Krein's
proposition, that relies on an integral criterion for the measure itself, in
order to show indeterminacy. In fact, taking into account the two
propositions and observing that $\int_{-\infty }^{\infty }x^{2n}\exp
(-\left\vert x\right\vert ^{\alpha })dx=2\alpha ^{-1}\Gamma \left( \frac{2n+1%
}{\alpha }\right) \simeq \left( \frac{2n}{\alpha }\right) !$ and that
Krein's proposition holds for $\exp \left( -\left\vert x\right\vert ^{\alpha
}\right) $ if $\alpha <1$, then we see that there are examples of Hamburger
indeterminate moment problems with growth just slightly faster than the $n!$
growth (the determinate case). The same applies for the Stieltjes case but
with $\alpha <1/2$ and the ($2n)!$ factorial growth.

It is explicitly stated in \cite{Simon} and wrongly assumed, for example in 
\cite{Kar}, that from the first proposition one might hope that just as the
condition of not too great growth implies determinacy, there might be a
condition of rapid growth that implies indeterminacy. This is false, since
there are moments of essentially arbitrary rates of growth which lead to
determinate moment problems. This is important both from the theoretical and
from the practical point of view. In fact, it can be related to an analogous
situation in Quantum Field Theory, when studying the role of power
expansions in QFT (perturbative QFT) and can be, to begin with, related with
resummation \cite{QFT2}\cite{QFT4}. We will comment on this later.

Given a set of moments $\left\{ \gamma _{n}\right\} _{n=0}^{\infty }$ and $%
c\in R$, one can define a new set of moments: 
\begin{equation}
\gamma _{n}(c)=\sum_{j=0}^{\infty } 
\begin{array}{l}
n \\ 
j%
\end{array}
c^{j}\gamma _{n-j},
\end{equation}
For the Hamburger problem, the solutions of the $\left\{ \gamma _{n}\right\}
_{n=0}^{\infty }$ and each of the $\left\{ \gamma _{n}(c)\right\}
_{n=0}^{\infty }$ are in one-to-one correspondence. The Stieltjes case is
much more involved in this respect, due essentially to the fact that its
associated interval $\left( 0,\infty \right) $ gets modified under the
change to a new set of moments (see \cite{Simon} for details). Another
well-known and useful condition on the moments is the Carleman criterion,
which states that: (i) if 
\begin{equation*}
\sum_{n=1}^{\infty }\gamma _{2n}^{-1/2n}=\infty ,
\end{equation*}
then the Hamburger problem is determinate, and (ii) if: 
\begin{equation*}
\sum_{n=1}^{\infty }\gamma _{n}^{-1/2n}=\infty ,
\end{equation*}
for a set of Stieltjes moments, that problem is both Stieltjes and Hamburger
determinate.

These few results already convey the idea that the classical moment problem,
in spite of the simplicity of its formulation (common to many inverse
problems) is highly nontrivial and deserves careful study. One important
aspect that we have not dealt with here is its tight relationship with the
theory of orthogonal polynomials. Since orthogonal polynomials can be
constructed from the set of moments, many of the conditions can be stated in
terms of the associated orthogonal polynomials, and these exhibit profound
mathematical properties depending on whether they come from a determinate or
from an undetermined measure.

The kind of mathematical concepts and arguments used here, as well as the
information presented regarding the theory of the classical moment problem
is clearly reminiscent to a fundamental problem in Quantum Field Theory,
namely the role of the perturbative expansion and its relationship with the
full theory. A huge amount of work has been devoted to clarify this issue,
since the early criticisms and observations on the drawbacks of the
perturbative expansion \cite{QFT1}. The literature on this problems is
extense, and we just point out here some conceptual remarks that can appear
in a natural way in the context of our problem. In view of the probable
divergence of QFT, the resummation of the perturbation series is necessary
for obtaining finite answers to physical problems. It is believed that
divergent expansions probably constitute asymptotic series, but the main
point is that it is not yet well known if a \textit{unique} answer is
implied by the perturbation theory. This is a fundamental issue. Namely, the
main question is not that of the convergence or divergence of the series,
but whether the expansion \textit{uniquely} determines the answer or not.
This is, of course, the aim of the renormalization program. There, one
source of problems, at the level of resummation, are the infrared
renormalons that are the responsible, for example, for the non-summability
of QED or QCD (see \cite{QFT2}\cite{QFT4} for details and references to the
literature). A point to be stressed is that the resummation of divergent
series can be especially ambiguous if the Carleman theorem, which guarantees
that there is a one-to-one correspondence between a function an its
associated asymptotic series, is not satisfied (see \cite{QFT4} and
references therein).

\section{Acknowledgements}

One of us (M.T.) warmly thanks Drs. P. Talavera and A.M. Turiel for a
careful reading of a preliminary version of the manuscript as well as Prof.
D. Kreimer for constant encouragement and support. This investigation has
been partly supported by DGICYT (Spain), project BFM2000-0810 and by CIRIT
(Generalitat de Catalunya), grant 1999SGR-00257.


\begin{thebibliography}{99}
\bibitem{Eli1} E. Elizalde \textit{Ten Physical application of spectral zeta
functions}. Lecture notes in Physics m 35 (Springer Verlag, Berlin, 1995).
E. Elizalde, S.D. Odintsov, A. Romeo, A.A. Bytsenko, and S. Zerbini, \textit{%
Zeta Regularization Techniques with Applications} (World Scientific,
Singapore, 1994). E. Elizalde, Commun. Math. Phys. \textbf{198} (1998) 83.
E. Elizalde, J. Comput. Appl. Math. \textbf{118} (2000) 125.

\bibitem{Eli2} E. Elizalde, L. Vanzo and S. Zerbini, Commun. Math. Phys. 
\textbf{194} (1998) 613; M. Bordag, E. Elizalde and K. Kirsten, J. Math.
Phys. \textbf{37} (1996) 895.

\bibitem{Simon} B. Simon, \textit{The classical moment problem as a
self-adjoint finite difference operator}. Advances in Mathematics. Vol 137
No 1. Jul 1998. pp 82-203

\bibitem{Eli3} E. Elizalde, J. Phys. \textbf{A22} (1989) 931.

\bibitem{QFT1} F.J. Dyson, Phys. Rev \textbf{85} (1952) 631; P.M. Stevenson,
Phys. Rev \textbf{D23} (1981); P.M. Stevenson, Nucl. Phys. \textbf{B231}
(1984) 65.

\bibitem{QFT2} J. Fischer, Int. J. Mod. Phys. \textbf{A12} (1997) 3625.

\bibitem{QFT3} S.A. Pernice and G. Oleaga, Phys. Rev. \textbf{D57} (1998)
1144; J. Zinn-Justin, \textit{Quantum Field Theory and Critical Phenomena'}.
Third Edition (Clarendon Press, Oxford, 1996).

\bibitem{QFT4} U.D. Jentschura, E.J. Weniger, and G. Soff, J. Phys. \textbf{%
G26} (2000) 1545.

\bibitem{Asympt} N. Bleistein and R.A. Handelsman, \textit{Asymptotic
Expansions of Integrals} (Dover, New York, 1986).

\bibitem{NC1} A. Connes, \textit{Noncommutative geometry} (Academic Press,
New York, 1994).

\bibitem{NC2} A. Connes, Geom. Func. Anal. (2000) 481.

\bibitem{Titch} E.C. Titchmarsh, \textit{The theory of the Riemann
Zeta-function} (Oxford, UK, 1986).

\bibitem{Gilkey} P. Gilkey, \textit{Invariance theory, the heat equation,
and the Atiyah-Singer Index Theorem}. 

\bibitem{NC3} Jose M. Gracia-Bondia, J. Varilly, and H. Figueroa, \textit{%
Elements of Noncommutative Geometry} (Birkhauser. Boston, 2000).

\bibitem{Kar} M. Kardar, in \textit{Fluctuating geometries in statistical
mechanics and field theory}. F. David, P. Ginsparg, and J. Zinn-Justin, Eds.
(Elsevier, Amsterdam, 1996); T. Emig and M. Kardar, cond-mat/0101247.

\bibitem{Eli4} E. Elizalde, J. Phys. \textbf{A34} (2001) 3025.

\bibitem{bez1} A.A. Bytsenko, E. Elizalde, and S. Zerbini, Phys. Rev. 
\textbf{D64} (2001) 105024.

\bibitem{fpw1} H. Falomir, P.A.G. Pisani, and A. Wipf, \textit{Pole structre
of the Hamiltonian $\zeta -$function for a singular potential},
math-ph/0112019.
\end{thebibliography}
\end{document}